\documentclass[]{svjour2}

\smartqed

\usepackage{amssymb}
\usepackage{graphicx}

\journalname{J Stat Phys}

\begin{document}

\title{Amplitude Function of Asymptotic Correlations Along Charged Wall
in Coulomb Fluids}

\titlerunning{Coulomb Fluids Near a Charged Wall}

\author{Ladislav \v{S}amaj}

\institute{Institute of Physics, Slovak Academy of Sciences, 
D\'ubravsk\'a cesta 9, SK-84511 Bratislava, Slovakia \\
\email{Ladislav.Samaj@savba.sk}}

\date{Received:  / Accepted: }

\maketitle

\begin{abstract}
In classical semi-infinite Coulomb fluids, two-point correlation 
functions exhibit a slow inverse-power law decay along a uniformly 
charged wall. 
In this work, we concentrate on the corresponding amplitude function 
which depends on the distances of the two points from the wall. 
Recently [L. \v{S}amaj, J. Stat. Phys. {\bf 161}, 227 (2015)], 
applying a technique of anticommuting variables to a 2D system of charged
rectilinear wall with ``counter-ions only'', we derived a relation between 
the amplitude function and the density profile which holds for any temperature.
In this paper, using the M\"obius conformal transformation of particle 
coordinates in a disc, a new relation between the amplitude function and 
the density profile is found for that model. 
This enables us to prove, at any temperature, the factorization property 
of the amplitude function in point distances from the wall and to express
it in terms of the density profile.
Presupposing the factorization property of the amplitude function and
using specific sum rules for semi-infinite geometries, a relation between
the amplitude function of the charge-charge structure function and the charge
profile is derived for many-component Coulomb fluids in any dimension.

\keywords{Coulomb fluid\and Counter-ions\and Free-fermion point \and Sum rules}

\end{abstract}

\renewcommand{\theequation}{1.\arabic{equation}}
\setcounter{equation}{0}

\section{Introduction} \label{Sect.1}
The topic of interest in this paper is the equilibrium statistical mechanics of 
classical Coulomb fluids which consist in mobile charges and perhaps fixed 
surface or volume charge densities, the system as a whole being electroneutral.
The charged entities interact by the Coulomb potential whose form
depends on the manifold in which the system is formulated.
For an infinite $d$-dimensional Euclidean space, the electrostatic 
potential $v$ at a point ${\bf r}\in \mathbb{R}^{d}$, induced by a unit charge 
at the origin ${\bf 0}$, is the solution of the Poisson equation
\begin{equation} \label{Poisson}
\Delta v({\bf r}) = - s_d \delta({\bf r}) ,
\end{equation}
where $s_d$ is the surface area of the unit sphere in $d$ dimensions:
$s_2=2\pi$, $s_3=4\pi$, etc.
This $d$-dimensional definition of the Coulomb potential maintains
generic properties of ``real'' 3D Coulomb systems with 
$v({\bf r})=1/r$, $r=\vert {\bf r}\vert$.
In 2D, the solution of (\ref{Poisson}), subject to the boundary 
condition $\nabla v({\bf r})\to 0$ as $r\to\infty$, reads as
$v({\bf r}) = - \ln(r/L)$ where the scale $L$ is free.
For $d\ge 3$, we have $v({\bf r})=1/r^{d-2}$.

In standard ``dense'' Coulomb fluids like the one-component plasma 
(jellium with a neutralizing bulk background) and the two-component plasma 
(Coulomb gas of $\pm$ charges), the number of mobile charges is proportional 
to the volume of the confining domain.
Such systems exhibit good screening properties and their bulk two-point
correlations have a short-ranged decay.
There exist many exact sum rules which relate the particle one-body
and two-body densities, in the bulk, semi-infinite and finite geometries,
see review \cite{Martin88}.

In ``sparse'' Coulomb systems of charged macromolecule surfaces, 
the number of identical mobile charges (coined as counter-ions) 
is proportional to the charged surface boundary from which 
they are released \cite{Attard96,Hansen00,Messina09}.  
The high-temperature (weak-coupling) limit is described by the
Poisson-Boltzmann (PB) theory \cite{Andelman} and by its systematic
improvement via the loop expansion \cite{Attard88,Netz00,Podgornik90}.
The low-temperature (strong-coupling) limit is more controversial,
the single-particle picture of counter-ions in the linear surface-charge 
potential appears in the leading strong-coupling order,
see e.g. \cite{Boroudjerdi05,Mallarino13,Mallarino15,Netz01,Samaj11a}.
In spite of the fact that sparse Coulomb fluids have poor screening
properties, the standard sum rules hold in semi-infinite and finite geometries
\cite{Samaj13,Samaj14,Samaj15}.  

In semi-infinite geometry of an electric double layer, the screening cloud 
around a particle sitting near hard wall is asymmetric and therefore 
two-point correlations decay slowly as an inverse-power law along the wall 
\cite{Jancovici82a,Jancovici82b,Usenko79}.
The corresponding amplitude function, which depends on the distances of
the two points from the wall, satisfies a sum rule
\cite{Jancovici82b,Jancovici95,Samaj10}. 
A relation between the amplitude function and the dipole moment was
found in Ref. \cite{Jancovici01}.
The contribution of the long-ranged charge-charge correlations along
a domain boundary, together with a bulk contribution, explains 
the dependence of the dielectric susceptibility tensor on the shape of 
the confining domain, in the thermodynamic limit, as required by macroscopic 
electrostatics \cite{Choquard86,Choquard87,Choquard89}.
Interestingly, in all exactly solvable cases the amplitude function
factorizes itself in the two point distances from the wall.

In a recent paper about 2D charged rectilinear wall with counter-ions only 
\cite{Samaj15}, we used a technique of anticommuting variables \cite{Samaj95} 
to derive a relation between the amplitude function and the density profile 
which holds for any coupling (temperature) of the fluid regime.
Moreover, using the M\"obius conformal transformation of particle coordinates 
in the partition function for a disc geometry, an exact formula for 
the dielectric susceptibility tensor was derived.
Since this tensor contains also long-ranged correlations along the wall,
it is likely that a more detailed exploration of the M\"obius conformal 
transformation might reveal another relation between the amplitude function 
and the density profile which is complementary to the one derived in
Ref. \cite{Samaj15}.

In this paper, using the formalism of anticommuting variables, we repeat 
the M\"obius conformal transformation of particle coordinates on the level 
of the partition function and one-body density (Sect. 2). 
In this way, we derive a new relation between the amplitude function and 
the density profile (Sect. 3).
This enables us to prove the factorization property of the amplitude function
for any temperature, at least for the simplified 2D model of the charged line 
with counter-ions only. 
The amplitude function is subsequently expressed locally in terms of the
density profile.
In Sect. 4, a relation of our result to the sum rule obtained by 
Blum et al. \cite{Blum81} enables us to extend the analysis to 
one-component jellium.
The generalization of the formalism to charge-charge structure function of
many-component Coulomb fluids in any dimension is presented in Sect. 5.
Here, presupposing the factorization property of the amplitude function,
its explicit relation to the charge density profile is established.  
A short recapitulation and conclusions are drawn in Sect. 6.

\renewcommand{\theequation}{2.\arabic{equation}}
\setcounter{equation}{0}

\section{2D charged rectilinear wall with counter-ions only}
We consider a system of $N$ identical pointlike particles of elementary
charge $-e$ confined to a 2D domain $D$ of points ${\bf r}=(x,y)$.
The system is studied within the canonical ensemble at the inverse
temperature $\beta=1/(k_{\rm B}T)$.  
The particle interaction part of the energy reads
$- e^2 \sum_{(i<j)=1}^N \ln\vert{\bf r}_i-{\bf r}_j\vert$,
where the free length scale $L$ is set to unity.
The one-body Boltzmann factor $w({\bf r}) = \exp[-\beta u({\bf r})]$
involves all external potentials (e.g. due to a neutralizing bulk or
surface background) acting on particles.
Introducing the coupling constant $\Gamma\equiv 2\gamma=\beta e^2$, 
the partition function is given by 
\begin{equation} \label{part}
Z_N(\gamma) = \frac{1}{N!} \int_D \prod_{i=1}^N 
\left[ {\rm d}^2 r_i\, w({\bf r}_i) \right] 
\prod_{(i<j)=1}^N \vert {\bf r}_i-{\bf r}_j\vert^{2\gamma} ,
\end{equation}
where we omit irrelevant constant prefactors.

The one-body density of particles at point ${\bf r}\in D$ is defined by
\begin{equation}
n({\bf r}) = \langle \hat{n}({\bf r}) \rangle , \qquad
\hat{n}({\bf r}) = \sum_{i=1}^N \delta({\bf r}-{\bf r}_i) ,
\end{equation}
where $\hat{n}({\bf r})$ is the microscopic density of particles and
$\langle \cdots \rangle$ denotes the statistical average over
canonical ensemble.
The corresponding averaged charge density is simply 
$\rho({\bf r}) = -e n({\bf r})$.
At two-particle level, one introduces the two-body densities
\begin{equation}
n_2({\bf r},{\bf r}') = \left\langle \sum_{(i\ne j)=1}^N 
\delta({\bf r}-{\bf r}_i) \delta({\bf r}'-{\bf r}_j) \right\rangle .  
\end{equation}
The one-body and two-body densities can be obtained from the partition
function (\ref{part}) in the standard way as the functional derivatives:
\begin{eqnarray}
n({\bf r}) & = & w({\bf r}) \frac{1}{Z_N} \frac{\delta Z_N}{\delta w({\bf r})},
\\ n_2({\bf r},{\bf r}') & = & w({\bf r})  w({\bf r}') \frac{1}{Z_N} 
\frac{\delta^2 Z_N}{\delta w({\bf r}) \delta w({\bf r}')} .
\end{eqnarray}

The two-body densities $n_2({\bf r},{\bf r}')$ decouple to the product
of densities $n({\bf r})$ and $n({\bf r}')$ at asymptotically large
distances $\vert {\bf r}-{\bf r}'\vert\to\infty$.   
Therefore it is useful to introduce the (truncated) Ursell functions
\begin{equation} \label{Ursell}
U({\bf r},{\bf r}') = n_2({\bf r},{\bf r'}) - n({\bf r}) n({\bf r}') 
\end{equation}
which vanish at $\vert {\bf r}-{\bf r}'\vert\to\infty$.
For one-component systems of particles of charge $-e$, the charge-charge 
structure function is defined as
\begin{eqnarray}
S({\bf r},{\bf r}') & = & e^2 \left[ 
\langle \hat{n}({\bf r}) \hat{n}({\bf r}') \rangle 
- n({\bf r}) n({\bf r}') \right] \nonumber \\ & = &  
e^2 \left[ U({\bf r},{\bf r}') + n({\bf r}) \delta({\bf r}-{\bf r}') \right] .
\end{eqnarray}
The structure and Ursell functions differ from one another by a term which
is nonzero only if the two points merge, i.e.
\begin{equation} \label{USasymp}
U({\bf r},{\bf r}') = \frac{S({\bf r},{\bf r}')}{e^2} \qquad
\mbox{if ${\bf r}\ne {\bf r}'$.} 
\end{equation}
For any finite or infinite domain $D$, the structure function satisfies 
the zeroth-moment sum rule \cite{Martin88}
\begin{equation} \label{zeroth}
\int_D {\rm d}^2 r\, S({\bf r},{\bf r}') =
\int_D {\rm d}^2 r'\, S({\bf r},{\bf r}') = 0 .
\end{equation}

\subsection{Formalism of anticommuting variables}
The formalism of anticommuting variables for 2D one-component plasmas
has been introduced in Ref. \cite{Samaj95} and developed further in 
Refs. \cite{Samaj00,Samaj04,Samaj11b,Samaj13,Samaj14}.
For $\gamma$ a positive integer, the partition function (\ref{part}) can be 
expressed in terms of two sets of anticommuting variables 
$\{ \xi_i^{(\alpha)},\psi_i^{(\alpha)} \}$ each with $\gamma$ components 
$(\alpha=1,\ldots,\gamma)$, defined on a discrete chain of $N$ sites 
$i=0,1,\ldots,N-1$, as follows
\begin{equation} \label{antipart}
Z_N(\gamma) = \int {\cal D}\psi {\cal D}\xi\, {\rm e}^{{\cal S}(\xi,\psi)} , 
\qquad {\cal S}(\xi,\psi) = \sum_{i,j=0}^{\gamma(N-1)} \Xi_i w_{ij} \Psi_j .
\end{equation}
Here, ${\cal D}\psi {\cal D}\xi \equiv \prod_{i=0}^{N-1} {\rm d}\psi_i^{(\gamma)}
\cdots {\rm d}\psi_i^{(1)} {\rm d}\xi_i^{(\gamma)} \cdots {\rm d}\xi_i^{(1)}$
and the action ${\cal S}(\xi,\psi)$ involves pair interactions of composite
operators
\begin{equation} \label{composite}
\Xi_i = \sum_{i_1,\ldots,i_{\gamma}=0\atop (i_1+\cdots+i_{\gamma}=i)}^{N-1}
\xi_{i_1}^{(1)} \cdots \xi_{i_{\gamma}}^{(\gamma)} , \qquad
\Psi_i = \sum_{i_1,\ldots,i_{\gamma}=0\atop (i_1+\cdots+i_{\gamma}=i)}^{N-1}
\psi_{i_1}^{(1)} \cdots \psi_{i_{\gamma}}^{(\gamma)} ,
\end{equation} 
i.e. the products of $\gamma$ anticommuting variables with the fixed
sum of site indices.
Using complex variables $z=x+{\rm i}y$ and $\bar{z}=x-{\rm i}y$, 
the interaction matrix is given by
\begin{equation} 
w_{ij} = \int_D {\rm d}^2 z\, z^i \bar{z}^j w(z,\bar{z}) , 
\qquad i,j=0,1,\ldots,\gamma(N-1) .
\end{equation}
The one-body and two-body densities are expressible explicitly as
\begin{eqnarray}
n({\bf r}) & = & w(z,\bar{z}) \sum_{i,j=0}^{\gamma(N-1)} 
\langle \Xi_i \Psi_j \rangle z^i \bar{z}^j , \\
n_2({\bf r}_1,{\bf r}_2) & = & w(z_1,\bar{z}_1) w(z_2,\bar{z}_2) 
\sum_{i_1,j_1,i_2,j_2=0}^{\gamma(N-1)} 
\langle \Xi_{i_1} \Psi_{j_1} \Xi_{i_2} \Psi_{j_2} \rangle
z_1^{i_1} \bar{z}_1^{j_1} z_2^{i_2} \bar{z}_2^{j_2} , \phantom{aaa}
\end{eqnarray}
where $\langle \cdots\rangle \equiv \int {\cal D}\psi {\cal D}\xi\, 
{\rm e}^S \cdots/Z_N(\gamma)$ 
denotes averaging over the anticommuting variables.

Next we consider the disc domain $D=\{ {\bf r}, \vert {\bf r} \vert \le R \}$
with a constant line charge density $\sigma e$ on the disc circumference $r=R$.
The requirement of the electroneutrality fixes the number of counter-ions
with charge $-e$ to $N=2\pi R\sigma$.
For this model, we have $w(z,\bar{z})\equiv w(r)=1$ \cite{Samaj15} and
\begin{equation} \label{wi}
w_{ij} = w_i\delta_{ij} , \qquad
w_i = 2\pi \int_0^R {\rm d}r\, r^{2i+1} = \frac{\pi}{i+1} R^{2(i+1)} .
\end{equation}
The diagonalization of the action in composite operators
\begin{equation} \label{S0}
{\cal S}(\xi,\psi) = \sum_{i=0}^{\gamma(N-1)} \Xi_i w_i \Psi_i 
\end{equation}
implies that 
$\langle \Xi_i\Psi_j \rangle = \delta_{ij} \langle \Xi_i\Psi_i \rangle$,
$\langle \Xi_{i_1}\Psi_{j_1} \Xi_{i_2}\Psi_{j_2}\rangle \ne 0$ only if
$i_1+j_1=i_2+j_2$, etc.
This fact simplifies the series representations of the one-body and two-body
densities:
\begin{eqnarray}
n(r) & = & \sum_{i=0}^{\gamma(N-1)} \langle \Xi_i \Psi_i \rangle
r^{2i} , \label{antione} \\
n_2({\bf r}_1,{\bf r}_2) & = & \sum_{i_1,j_1,i_2,j_2=0\atop (i_1+i_2=j_1+j_2)}^{\gamma(N-1)} 
\langle \Xi_{i_1} \Psi_{j_1} \Xi_{i_2} \Psi_{j_2} \rangle
z_1^{i_1} \bar{z}_1^{j_1} z_2^{i_2} \bar{z}_2^{j_2} . \label{antitwo}
\end{eqnarray}

\subsection{Conformal transformation}
We consider the particles with complex coordinates $(z,\bar{z})$
inside the disc domain $D=\{ (z,\bar{z}), z\bar{z} \le R^2 \}$. 
The M\"obius conformal transformation
\begin{equation} \label{conftrans}
z' = \frac{z + R a}{1+\frac{z\bar{a}}{R}} , \qquad
z = \frac{z'- R a}{1-\frac{z'\bar{a}}{R}} 
\end{equation}
(with a free complex parameter $a$ such that $a\bar{a} \ne 1$)
transforms the particle coordinates in the disc domain $D$ to another
domain $D'$ defined by the inequality
\begin{equation}
( R^2 - z'\bar{z}') (1-a\bar{a}) \ge 0 .
\end{equation}
If $a$ is chosen such that $a \bar{a} < 1$, the original disc domain $D$ 
is mapped onto itself, $D'=D$.
Note that $a=0$ corresponds to the identity transformation.

\subsubsection{Partition function}
Let us study the effect of the M\"obius transformation of all particle
coordinates on the partition function
\begin{equation} \label{part1}
Z_N(\gamma) = \frac{1}{N!} \int_D \prod_{i=1}^N {\rm d} z_i {\rm d} 
\bar{z}_i\, \prod_{(i<j)=1}^N \vert z_i-z_j\vert^{2\gamma} .
\end{equation}
Under the conformal transformation (\ref{conftrans}), each surface element 
${\rm d}z{\rm d}\bar{z}$ transforms as
\begin{equation}
{\rm d}z {\rm d}\bar{z} = \frac{(1-a\bar{a})^2}{\left( 1-\frac{z'\bar{a}}{R} 
\right)^2 \left( 1-\frac{\bar{z}'a}{R} \right)^2} {\rm d}z' {\rm d}\bar{z}' 
\end{equation}
and each square of the distance between two particles transforms as
\begin{equation}
\vert z_i-z_j\vert^2 = \frac{(1-a\bar{a})^2}{\left( 1-\frac{z'_i\bar{a}}{R} 
\right) \left( 1-\frac{\bar{z}'_i a}{R} \right) \left( 1-\frac{z'_j\bar{a}}{R} 
\right) \left( 1-\frac{\bar{z}'_j a}{R} \right)} \vert z'_i-z'_j\vert^2 .
\end{equation}
The partition function (\ref{part1}) can be written in terms of 
the transformed coordinates as follows
\begin{equation} \label{part2}
Z_N^a(\gamma) = \frac{1}{N!} \int_D \prod_{i=1}^N {\rm d} z'_i {\rm d} 
\bar{z}'_i \left[ \frac{(1-a\bar{a})}{\left( 1-\frac{z'_i\bar{a}}{R} \right) 
\left( 1-\frac{\bar{z}'_i a}{R} \right)} \right]^{\nu}
\prod_{(i<j)=1}^N \vert z'_i-z'_j\vert^{2\gamma} ,
\end{equation}
where we use the notation $\nu\equiv \gamma(N-1)+2$. 
The transformed variables $z'$ and $\bar{z}'$ under integration can be 
replaced by the original ones $z$ and $\bar{z}$.
We see that the effect of the conformal transformation consists in changing 
the circular one-body Boltzmann factor $w(r) = 1$ to the non-circular one
\begin{equation}
w^a(z,\bar{z}) = 
\left[ \frac{(1-a\bar{a})}{\left( 1-\frac{z\bar{a}}{R} \right) 
\left( 1-\frac{\bar{z} a}{R} \right)} \right]^{\nu} .
\end{equation}
The diagonal ${\cal S}$-action (\ref{S0}) transforms itself into 
the non-diagonal one
\begin{equation} \label{Sa}
{\cal S}^a(\xi,\psi) = \sum_{i,j=0}^{\gamma(N-1)} \Xi_i w_{ij}^a \Psi_j , 
\end{equation}
where
\begin{equation} \label{wij}
w_{ij}^a = \int_D {\rm d}^2 z\, 
\left[ \frac{(1-a\bar{a})}{\left( 1-\frac{z\bar{a}}{R} \right) 
\left( 1-\frac{\bar{z} a}{R} \right)} \right]^{\nu} z^i \bar{z}^j , 
\qquad i,j=0,1,\ldots,\gamma(N-1) .
\end{equation}
The equivalence of the original partition function $Z_N(\gamma;\{ w_i\})$
with the transformed one $Z_N^a(\gamma;\{ w_{ij}^a\})$,
\begin{equation} \label{Z1}
Z_N(\gamma) = Z_N^a(\gamma) , 
\end{equation}
can be expressed in terms of the integrals over anticommuting variables as
\begin{equation} \label{Z2}
\int {\cal D}\psi {\cal D}\xi\, \exp\left[ {\cal S}(\xi,\psi) \right] =
\int {\cal D}\psi {\cal D}\xi\, \exp\left[ {\cal S}^a(\xi,\psi) \right] .  
\end{equation}

\subsubsection{Particle density}
Under the conformal transformation (\ref{conftrans}), the density 
$n(z,\bar{z};\{ w_i\}) \equiv n(r)$ transforms itself to 
$n^a(z',\bar{z}';\{ w_{ij}^a\})$ according to
\begin{equation}
n(z,\bar{z}) {\rm d}z {\rm d}\bar{z} = n^a(z',\bar{z}') 
{\rm d}z' {\rm d}\bar{z}' .
\end{equation}
Note that this relation, when integrated over the disk domain $D$, ensures
the conservation of the total number of particles under the conformal 
transformation.
Equivalently,
\begin{equation}
n(z,\bar{z}) = \left[ \frac{\left( 1-\frac{z'\bar{a}}{R} \right) 
\left( 1-\frac{\bar{z}' a}{R} \right)}{(1-a\bar{a})} \right]^2
n^a(z',\bar{z}') .
\end{equation}
Within the formalism of anticommuting variables, the transformed particle 
density is expressible as
\begin{equation}
n^a(z',\bar{z}') = w^a(z',\bar{z}') \sum_{i,j=0}^{\gamma(N-1)} 
\langle \Xi_i\Psi_j \rangle^a (z')^i (\bar{z}')^j ,
\end{equation}
where the symbol $\langle \cdots\rangle^a$ means the averaging with
the ${\cal S}^a$-action (\ref{Sa}).
We conclude that
\begin{equation} \label{nr}
n(r) = \left[ \frac{(1-a\bar{a})}{\left( 1-\frac{z'\bar{a}}{R} \right) 
\left( 1-\frac{\bar{z}' a}{R} \right)} \right]^{\gamma(N-1)}  
\sum_{i,j=0}^{\gamma(N-1)} \langle \Xi_i\Psi_j \rangle^a (z')^i (\bar{z}')^j .
\end{equation}

\renewcommand{\theequation}{3.\arabic{equation}}
\setcounter{equation}{0}

\section{Derivation of sum rules}
In this part, we use the above exact relations between the original and
transformed partition functions and particle densities to derive 
certain sum rules.

\subsection{Partition function}
We start with the equality of the original and transformed partition
functions, see Eqs. (\ref{Z1}) and (\ref{Z2}).
First we expand the transformed interaction matrix (\ref{wij}) in
linear $a$, $\bar{a}$ and quadratic $a\bar{a}$ terms:
\begin{equation}
w_{ij}^a = \delta_{ij} w_i + \frac{\nu a}{R} \delta_{i,j+1} w_i
+ \frac{\nu \bar{a}}{R} \delta_{i+1,j} w_{i+1} 
+ a \bar{a} \delta_{ij} \left( \frac{\nu^2}{R^2} w_{i+1} -
\nu w_i \right) + \cdots , 
\end{equation}
where $w_i$ are the original interaction strengths (\ref{wi}).
The corresponding expansion of the transformed action (\ref{Sa}) 
around the original action (\ref{S0}) reads as
\begin{eqnarray}
{\cal S}^a & = & {\cal S} + \frac{\nu a}{R} \sum_i \Xi_{i+1} w_{i+1} \Psi_i
+ \frac{\nu \bar{a}}{R} \sum_i \Xi_i w_{i+1} \Psi_{i+1} \nonumber \\
& & + a \bar{a} \sum_i \Xi_i \left( \frac{\nu^2}{R^2} w_{i+1} -
\nu w_i \right) \Psi_i + \cdots . \label{Saexp}
\end{eqnarray} 
Inserting this expansion into Eq. (\ref{Z2}) and expanding the exponential
in $a$, $\bar{a}$ and $a\bar{a}$ terms, we obtain
\begin{eqnarray}
Z_N(\gamma) & = & Z_N(\gamma) \Bigg[ 1 + 
a \bar{a} \sum_i \langle \Xi_i \Psi_i \rangle 
\left( \frac{\nu^2}{R^2} w_{i+1} - \nu w_i \right) \nonumber \\ & &
+ a \bar{a} \frac{\nu^2}{R^2} \sum_{i,j} w_{i+1} w_{j+1}
\langle \Xi_i \Psi_{i+1} \Xi_{j+1} \Psi_j \rangle + \cdots \Bigg] .
\end{eqnarray}
The term proportional to $a\bar{a}$ must vanish.
Simultaneously, there holds
\begin{equation}
\sum_i w_i \langle \Xi_i\Psi_i \rangle = \int_D {\rm d}^2r\, n(r) = N
\end{equation}
and
\begin{equation}
\sum_i w_{i+1} \langle \Xi_i\Psi_i \rangle = \int_D {\rm d}^2r\, r^2 n(r) .
\end{equation}
From the representation (\ref{antitwo}) we get
\begin{equation}
\sum_{i,j} w_{j+1} \langle \Xi_i \Psi_{i+1} \Xi_{j+1} \Psi_j \rangle r^{2(i+1)}
= \int_D {\rm d}^2r'\, {\bf r}\cdot {\bf r}' n_2({\bf r},{\bf r}') ,
\end{equation}
where ${\bf r}\cdot {\bf r}' = (z \bar{z}' + \bar{z} z')/2$ denotes
the scalar product of vectors ${\bf r}$ and ${\bf r}'$.
Consequently, we end up with the sum rule
\begin{equation} \label{sum1}
\int_D {\rm d}^2 r \int_D {\rm d}^2 r'\, {\bf r}\cdot {\bf r}'
\langle \hat{n}({\bf r}) \hat{n}({\bf r}') \rangle 
= \frac{R^2 N}{\gamma(N-1)+2} .
\end{equation}
This sum rule has already been derived in connection with the calculation of 
the dielectric susceptibility tensor, see Eq. (6.10) of Ref. \cite{Samaj15}.
The same equality holds for the truncated correlator
$\langle \hat{n}({\bf r}) \hat{n}({\bf r}') \rangle - n(r) n(r')$ since
\begin{equation}
\int_D {\rm d}^2r \int_D {\rm d}^2r'\, {\bf r}\cdot {\bf r}' n(r) n(r') = 0
\end{equation}
after the integration of $\cos(\varphi-\varphi')$ over the angle
$\varphi-\varphi'$ from $0$ to $2\pi$.

\subsection{Particle density}
In the density relation (\ref{nr}), we expand up to terms linear in $a$
and $\bar{a}$ the transformed coordinates
\begin{equation}
z' = z + Ra - \frac{z^2}{R} \bar{a} + \cdots , \qquad
\bar{z}' = \bar{z} + R\bar{a} - \frac{\bar{z}^2}{R} a + \cdots 
\end{equation}
and, with the aid of the $S^a$-expansion (\ref{Saexp}), 
the transformed correlators 
\begin{eqnarray}
\langle \Xi_i\Psi_j\rangle^a & = & \delta_{ij} \langle \Xi_i\Psi_i\rangle 
+ \delta_{i+1,j} \frac{\nu a}{R} \sum_k w_{k+1} 
\langle \Xi_i\Psi_{i+1}\Xi_{k+1}\Psi_k \rangle \nonumber \\ & &  
+ \delta_{i,j+1} \frac{\nu \bar{a}}{R} \sum_k w_{k+1} 
\langle \Xi_{i+1}\Psi_i\Xi_k\Psi_{k+1} \rangle + \cdots . 
\end{eqnarray}
Thus we obtain
\begin{eqnarray}
0 & = & \frac{\gamma(N-1)}{R} (\bar{z}a+z\bar{a}) n(r)
+ \frac{\nu a}{R} \sum_{ij} w_{j+1} 
\langle \Xi_i\Psi_{i+1}\Xi_{j+1}\Psi_j \rangle z^i \bar{z}^{i+1} \nonumber \\ & &
+ \frac{\nu \bar{a}}{R} \sum_{ij} w_{j+1} 
\langle \Xi_{i+1}\Psi_i\Xi_j\Psi_{j+1} \rangle z^{i+1} \bar{z}^i \nonumber \\ & &
+ (\bar{z}a+z\bar{a}) \sum_i \langle \Xi_i\Psi_i\rangle (z\bar{z})^i i
\left( \frac{R}{z\bar{z}} - \frac{1}{R} \right) .
\end{eqnarray}
After simple algebra, this result leads to the equality
\begin{equation} \label{sum2}
\left[ \gamma(N-1)+2 \right] \int_D {\rm d}^2 r'\, {\bf r}\cdot {\bf r}'
\langle \hat{n}({\bf r}) \hat{n}({\bf r}') \rangle 
= 2 r^2 n(r) - \frac{1}{2} r (R^2-r^2) \frac{\partial}{\partial r} n(r) .
\end{equation}
Applying the integration $\int_D {\rm d}^2 r$ to this relation, it reduces 
to the previous sum rule (\ref{sum1}), but the present sum rule is more 
informative.

Let the vector ${\bf r}$ be taken along the $x$-axis, ${\bf r}=(r,0)$,
so that ${\bf r}\cdot {\bf r}' = r r' \cos\varphi'$.
Substituting the correlator in (\ref{sum2}) by its truncation
$\langle \hat{n}({\bf r}) \hat{n}({\bf r}') \rangle - n(r) n(r') 
\equiv S({\bf r},{\bf r}')/e^2$ and using the zeroth-moment sum rule
(\ref{zeroth}), we obtain
\begin{eqnarray}
[\gamma(N-1)+2] \left[ \int_0^{2\pi} {\rm d}\varphi'
\int_0^R {\rm d}r'\, (r')^2 (1-\cos\varphi') \frac{S({\bf r},{\bf r}')}{e^2} 
\phantom{aaaaaaa} \right. \nonumber \\ 
\left. + \int_D {\rm d}^2r'\, (R-r') \frac{S({\bf r},{\bf r}')}{e^2} \right] 
= - 2 r n(r) + \frac{1}{2} (R-r) (R+r) \frac{\partial}{\partial r} n(r) .
\label{pom1}
\end{eqnarray}
To go from the disc to the semi-infinite rectilinear geometry in the limit 
$R\to\infty$, we switch to the variables $x = R - r$ and $x' = R - r'$.
Eq. (\ref{pom1}) then becomes
\begin{eqnarray}
\frac{[\gamma(N-1)+2]}{R} \left[ \frac{1}{2} \int_{-\pi}^{\pi} 
{\rm d}\varphi \int_0^{\infty} {\rm d}x'\,  
\left( 2 R \sin\frac{\varphi}{2} \right)^2 \frac{S(x,x';\varphi)}{e^2} 
\phantom{aaaaa} \right. \nonumber \\ 
\left. + \int_{-\pi}^{\pi} {\rm d}(R\varphi) \int_0^{\infty} {\rm d}x'\, x' 
\frac{S(x,x';\varphi)}{e^2} \right] = - \left[ 
x \frac{\partial}{\partial x} n(x) + 2 n(x) \right] . \label{eqno}
\end{eqnarray}

For the disc geometry it was shown \cite{Jancovici02} that, as the radius
$R\to\infty$, the Ursell function of two particles at finite distances
$x$ and $x'$ from the disc boundary and with the angle $\varphi\ne 0$
between them behaves as
\begin{equation} \label{asympdisc}
U(x,x';\varphi) = \frac{S(x,x';\varphi)}{e^2}
\mathop{\simeq}_{R\to\infty} \frac{f(x,x')}{[2R\sin(\varphi/2)]^2} , \qquad
\varphi\ne 0 .
\end{equation} 
Since $(x'-x)\delta(x-x')=0$, we can also write
\begin{eqnarray}
\int_0^{\infty} {\rm d}x'\, x' \frac{S(x,x';\varphi)}{e^2} & = & 
\int_0^{\infty} {\rm d}x'\, (x'-x) \frac{S(x,x';\varphi)}{e^2} \nonumber \\
& = & \int_0^{\infty} {\rm d}x'\, (x'-x) U(x,x';\varphi) .
\end{eqnarray}
Introducing the lateral distance $y = R \varphi$ for the rectilinear
geometry, Eq. (\ref{eqno}) becomes
\begin{eqnarray}
2\pi\gamma\sigma \left[ \pi \int_0^{\infty} {\rm d}x'\, f(x,x')
+ \int_{-\infty}^{\infty} {\rm d}y \int_0^{\infty} {\rm d}x'\, (x'-x)
U(x,x';y) \right] \nonumber \\ 
= - \left[ x \frac{\partial}{\partial x} n(x) + 2 n(x) \right] . 
\label{vys}
\end{eqnarray}
There exists a 2D relation between asymptotic behavior and dipole moment 
seen from a fixed point with coordinate $x$
\cite{Jancovici01}:
\begin{equation} \label{rov1}
\int_{-\infty}^{\infty} {\rm d}y \int_0^{\infty} {\rm d}x'\, (x'-x) U(x,x';y)
= \pi \int_0^{\infty} {\rm d}x'\, f(x,x') . 
\end{equation}
Consequently, Eq. (\ref{vys}) implies that 
\begin{equation} \label{rov2}
\int_0^{\infty} {\rm d}x'\, f(x,x') = - \frac{1}{2\pi^2\Gamma\sigma}
\left[ x \frac{\partial}{\partial x} n(x) + 2 n(x) \right] . 
\end{equation}

In the previous paper \cite{Samaj15}, we found that
\begin{equation} \label{rov3}
\pi f(x,0) =  - \left[ x \frac{\partial}{\partial x} n(x) + 2 n(x) \right] . 
\end{equation}
Combining the last two equations, we finally arrive at the relation
\begin{equation} \label{crucial}
\int_0^{\infty} {\rm d}x'\, f(x,x') = \frac{1}{2\pi\Gamma\sigma} f(x,0)
\end{equation}
containing only the function of interest $f(x,x')$.

This equation can be checked on two exactly solvable 2D cases of 
the present counter-ion system \cite{Samaj15}.
The PB $\Gamma\to 0$ limit yields $f(x,x')$ of the long-range form 
\cite{Samaj13}
\begin{equation} \label{PBcounter}
f(x,x') = - \frac{2}{\pi^2\Gamma} \frac{b^4}{(x+b)^3(x'+b)^3} , \qquad
b = \frac{1}{\pi\Gamma\sigma} . 
\end{equation}
At $\Gamma=2$, we have $f(x,x')$ of the short-range form 
\cite{Jancovici84,Samaj13}
\begin{equation} \label{Gamma2counter}
f(x,x') = - 4 \sigma^2 {\rm e}^{-4\pi\sigma x} {\rm e}^{-4\pi\sigma x'} .
\end{equation}
It is easy to verify that both exact solutions fulfill our Eq. (\ref{crucial}). 

\subsection{Properties of the amplitude $f$-function}
Now we aim at showing fundamental properties of the $f$-functions
following from Eq. (\ref{crucial}).

It is known that in 2D the function $f(x,x')$ obeys the sum rule
\cite{Jancovici82b,Jancovici95,Samaj10} 
\begin{equation} \label{sumrule2}
\int_0^{\infty} {\rm d}x \int_0^{\infty} {\rm d}x'\, f(x,x') 
= - \frac{1}{2\pi^2\Gamma} .
\end{equation}
Applying the integration operation $\int_0^{\infty} {\rm d}x$ to 
Eq. (\ref{crucial}), we get
\begin{equation}
\int_0^{\infty} {\rm d}x\, f(x,0) = - \frac{\sigma}{\pi} .
\end{equation}
Taking $x=0$ in (\ref{crucial}) and using the symmetry 
$f(x,x')=f(x',x)$, the $f$-function with both points at the $x=x'=0$ 
boundary is given by
\begin{equation}
f(0,0) = - 2\Gamma \sigma^2 .
\end{equation}
As a check, the exactly solvable $\Gamma\to 0$ limit (\ref{PBcounter}) and 
$\Gamma=2$ case (\ref{Gamma2counter}) satisfy this relation. 

The function $f(x,x')$ is assumed to be an analytic holomorphic function
of its arguments.
Therefore, when deriving both sides of the relation (\ref{crucial}) with 
respect to $x$, one can interchange the integration and derivation 
\cite{Tutschke} to obtain   
\begin{equation} \label{crucial1}
\int_0^{\infty} {\rm d}x'\, \frac{\partial f(x,x')}{\partial x} 
= \frac{1}{2\pi\Gamma\sigma} \frac{\partial f(x,0)}{\partial x} .
\end{equation}
The two Eqs. (\ref{crucial}) and (\ref{crucial1}) can be fulfilled 
simultaneously only if
\begin{equation}
\frac{\partial f(x,x')}{\partial x} = h(x) f(x,x') 
\end{equation}
with some unknown function $h(x)$.
Equivalently,
\begin{equation}
\frac{\partial}{\partial x} \ln\left[ -f(x,x') \right] = h(x) .
\end{equation}
Taking into account the symmetry relation $f(x,x') = f(x',x)$, this
PDE has the unique solution
\begin{equation}
\ln\left[ -f(x,x') \right] = \int {\rm d}x\, h(x)  + \int {\rm d}x'\, h(x') .
\end{equation}
Consequently, the function $f(x,x')$ factorizes as follows 
\begin{equation} \label{factor}
f(x,x') = - g(x) g(x') , \qquad g(x) = \exp\left[ \int {\rm d}x\, h(x) \right] .
\end{equation}
The factorization property of $f(x,x')$, seen in the $\Gamma\to 0$ limit 
(\ref{PBcounter}) and at $\Gamma=2$ (\ref{Gamma2counter}), thus extends
to any value of $\Gamma$.

Due to the factorization property, the density profile $n(x)$ determines
the function $f(x,x')$ as follows 
\begin{equation} \label{finalgen}
f(x,x') = - \frac{1}{2\pi^2\Gamma\sigma^2} 
\left[ x \frac{\partial n(x)}{\partial x} + 2 n(x) \right]
\left[ x' \frac{\partial n(x')}{\partial x'} + 2 n(x') \right] .  
\end{equation}
The prefactor is fixed by the sum rule (\ref{sumrule2}) together with
the equality
\begin{equation} \label{formula}
\int_0^{\infty} {\rm d}x\, \left[ 
x \frac{\partial n(x)}{\partial x} + 2 n(x) \right]
= \int_0^{\infty} {\rm d}x\, n(x) = \sigma ,
\end{equation}
where we used the integration by parts for $x\partial n(x)/\partial x$, 
the known fact that $n(x)$ goes to 0 faster than $1/x$ as $x\to\infty$ and 
the electroneutrality condition.

\renewcommand{\theequation}{4.\arabic{equation}}
\setcounter{equation}{0}

\section{Another approach to one-component systems}

\subsection{Counter-ions only}
There exists an alternative way how to derive in the 2D case with
counter-ions only the important relation (\ref{rov2}).
In 2D, the coupling constant $\Gamma=\beta e^2$ is dimensionless.
The particle density $n$ has dimension [length]$^{-2}$ and the
surface charge density $\sigma$ has dimension [length]$^{-1}$,
so one can write
\begin{equation}
n(x;\sigma) = \sigma^2 t(\sigma x) ,
\end{equation} 
where $t$ is an unknown function.
For this scaling form of the density profile, we obtain the equality
\begin{equation} \label{derns}
\sigma \frac{\partial n(x)}{\partial \sigma} 
= 2 \sigma^2 t(\sigma x) + \sigma^3 x t'(\sigma x)
= 2 n(x) + x \frac{\partial n(x)}{\partial x} .  
\end{equation}

Blum et al. \cite{Blum81} derived a sum rule which relates the variation
of the particle density $n(x)$ with respect to the surface charge density
to the dipole moment seen by a fixed particle.
In 2D, the sum rule reads as
\begin{equation} \label{Blum2D}
\frac{\partial n(x)}{\partial \sigma} = - 2\pi\Gamma
\int_{-\infty}^{\infty} {\rm d}y \int_0^{\infty} {\rm d}x'\, (x'-x) U(x,x';y)
\end{equation}
With the aid of the relations (\ref{rov1}) and (\ref{derns}), we recover
Eq. (\ref{rov2}).

We can go to higher dimensions $d$ within the present approach.
The $d$-dimensional Blum counterpart of the 2D sum rule (\ref{Blum2D}) 
is \cite{Blum81} 
\begin{equation} \label{Blumd}
\frac{\partial n(x)}{\partial \sigma} = - s_d \beta e^2
\int_{-\infty}^{\infty} {\rm d}y \int_0^{\infty} {\rm d}x'\, (x'-x) U(x,x';y) .
\end{equation}
The 2D relation between asymptotic behavior and dipole moment (\ref{rov1})
takes in $d$ dimensions the form \cite{Jancovici01}
\begin{equation} \label{Uf} 
\int_{-\infty}^{\infty} {\rm d}y \int_0^{\infty} {\rm d}x'\, (x'-x) U(x,x';y)
= \frac{s_d}{2} \int_0^{\infty} {\rm d}x'\, f(x,x')
\end{equation}
so that
\begin{equation} \label{rovnica}
\frac{\partial n(x)}{\partial \sigma} = - \frac{s_d^2}{2} \beta e^2
\int_0^{\infty} {\rm d}x'\, f(x,x') .
\end{equation}
This formula can be readily checked on the exactly solvable PB limit 
$\Gamma\to 0$ in any dimension $d$ \cite{Samaj13,Samaj15}:
\begin{equation} \label{exactd}
n(x) = \frac{\sigma b}{(x+b)^2} , \qquad
f(x,x') = - \frac{8}{\beta e^2 s_d^2} \frac{b^4}{(x+b)^3 (x'+b)^3} ,
\end{equation}
where $b=2/(\beta e^2\sigma s_d)$ is the Gouy-Chapmann length.

We cannot prove in general the factorization property (\ref{factor}) of 
the function $f(x,x')$ in dimensions $d\ge 3$ since we miss a relation like 
the 2D one (\ref{rov3}) derived for any temperature in Ref. \cite{Samaj15}.
Let us suppose that the factorization property takes place, i.e. 
$f(x,x')= - g(x) g(x')$, and apply the present formalism to obtain $g(x)$.
The generalization of the 2D sum rule (\ref{sumrule2}) to any dimension $d$
reads as
\begin{equation} \label{sumruled}
\int_0^{\infty} {\rm d}x \int_0^{\infty} {\rm d}x'\, f(x,x') 
= - \frac{2}{\beta e^2 s_d^2} .
\end{equation} 
Inserting the factorization assumption into this equation implies
\begin{equation} \label{r1}
\int_0^{\infty} {\rm d}x\, g(x) = \sqrt{\frac{2}{\beta e^2 s_d^2}} .
\end{equation}
Then, considering $f(x,x')= - g(x) g(x')$ in Eq. (\ref{rovnica}) leads to
\begin{equation} \label{r2}
g(x) = \sqrt{\frac{2}{\beta e^2 s_d^2}} \frac{\partial n(x)}{\partial \sigma} ,
\end{equation}
i.e. for every distance $x$ from the wall the function $g(x)$ is expressible 
locally in terms of the density profile.
Note that the relations (\ref{r1}) and (\ref{r2}) are fully consistent since
the integration of Eq. (\ref{r2}) over $x$ from $0$ to $\infty$ reduces 
to (\ref{r1}) due to the electroneutrality condition 
$\int_0^{\infty} {\rm d}x\, n(x) = \sigma$.
The factorized $d$-dimensional PB solution (\ref{exactd}) with 
\begin{equation}
g(x) = \sqrt{\frac{8}{\beta e^2 s_d^2}} \frac{b^2}{(x+b)^3}
\end{equation}
evidently fulfills Eq. (\ref{r2}). 

\subsection{Jellium model}
The relation (\ref{r2}) in fact holds for an arbitrary one-component system
whose $f(x,x')$-function factorizes into $- g(x) g(x')$.
Here, we present the jellium model of mobile pointlike particles
with charge $-e$ immersed in a homogeneous (bulk) background of density 
$n_0$ and charge density $e n_0$.
The system is constrained to the $d$-dimensional Euclidean half-space of points 
${\bf r}=(x,{\bf y})$ with ${\bf y}=(y_1,\ldots,y_{d-1})$, the coordinates  
$y_i\in (-\infty,\infty)$ and $x\ge 0$. 
There is a plane charged by a constant surface charge density $\sigma e$ 
at $x=0$. 
The density profile $n(x)$ and the function $f(x,x')$ were calculated
exactly in two cases.

The high-temperature Debye-H\"uckel (linearized PB) theory in $d$ dimensions
\cite{Jancovici82a} yields the density profile
\begin{equation}
n(x,\sigma) = n(x,\sigma=0) + \sigma \kappa {\rm e}^{-\kappa x} ,
\end{equation}
where $\kappa=\sqrt{s_d\beta e^2 n_0}$ is the inverse Debye length.
The asymptotic $\vert {\bf y}\vert = y \to\infty$ decay of 
the Ursell function along the wall was found in the form 
\begin{equation}
U(x,x';y) \mathop{\simeq}_{y\to\infty} - \frac{2 n_0}{s_d} {\rm e}^{-\kappa x}
{\rm e}^{-\kappa x'} \frac{1}{y^d} ,
\end{equation}
i.e.
\begin{equation}
g(x) = \sqrt{\frac{2 n_0}{s_d}} {\rm e}^{-\kappa x} .
\end{equation}
With regard to the equality
$\partial n(x)/\partial\sigma = \kappa {\rm e}^{-\kappa x}$, it is easy to 
verify that this $g$-function satisfies Eq. (\ref{r2}).

The other exactly solvable case is the 2D jellium at coupling $\Gamma=2$.
The free-fermion method \cite{Jancovici82a} yields the density profile
\begin{equation}
n(x,\sigma) = n_0 \frac{2}{\sqrt{\pi}} \int_{-\pi\sigma\sqrt{2}}^{\infty} {\rm d}t\,
\frac{1}{1+\phi(t)} {\rm e}^{-(t-x\sqrt{2})^2} ,
\end{equation}
where $\phi$ denotes the error function
\begin{equation}
\phi(t) = \frac{2}{\sqrt{\pi}} \int_0^t {\rm d}u\, {\rm e}^{-u^2} .
\end{equation}
The asymptotic decay of the Ursell function along the wall
\begin{equation}
U(x,x';y) \mathop{\simeq}_{y\to\infty} - n_0^2 \frac{2}{\pi} \frac{\exp\left\{ 
-2\left[ x^2+x'^2 + 2\pi\sigma(x+x') + 2\pi^2\sigma^2\right]\right\}}{
\left[ 1+\phi(-\pi\sigma\sqrt{2})\right]^2} \frac{1}{y^2} 
\end{equation}
implies the $g$-function of the form
\begin{equation}
g(x) = n_0 \sqrt{\frac{2}{\pi}} \frac{{\rm e}^{-2(x+\pi\sigma)^2}}{
\left[ 1+\phi(-\pi\sigma\sqrt{2})\right]} .
\end{equation}
After simple algebra it can be shown that
\begin{equation}
g(x) = \frac{1}{2\pi} \frac{\partial n(x)}{\partial \sigma} 
\end{equation}
which is in agreement with our result (\ref{r2}).

\renewcommand{\theequation}{5.\arabic{equation}}
\setcounter{equation}{0}

\section{A generalization to many-component Coulomb systems}
Now let us consider a general Coulomb system which consists of $s$ species
of particles $\alpha=1,\ldots,s$ with the corresponding charges $q_{\alpha} e$ 
($q_{\alpha}$ is the valence and $e$ the elementary charge), plus perhaps
a fixed background of density $n_0$ and charge density $\rho_0=e n_0$.
As before, the particles are constrained to the $d$-dimensional Euclidean 
half-space of points ${\bf r}=(x,{\bf y})$ with $x\ge 0$. 
There is a plane charged by a constant surface charge density $\sigma e$ 
at $x=0$.  
The microscopic density of particles of species $\alpha$ is given by
$\hat{n}_{\alpha}({\bf r}) = \sum_i \delta_{\alpha,\alpha_i} 
\delta({\bf r}-{\bf r}_i)$, where $i$ indexes the charged particles. 
The total microscopic charge density reads as $\hat{\rho}({\bf r}) 
= \rho_0 + \sum_{\alpha} q_{\alpha}e \hat{n}_{\alpha}({\bf r})$.
For the present geometry, the averaged charge density depends only on
the $x$-coordinate, $\rho(x) = \langle \hat{\rho}({\bf r}) \rangle$.
The charge-charge structure function, defined by
\begin{equation} \label{charge-charge}
S({\bf r},{\bf r}') \equiv 
\langle \hat{\rho}({\bf r}) \hat{\rho}({\bf r}') \rangle
- \langle \hat{\rho}({\bf r}) \rangle \langle \hat{\rho}({\bf r}') \rangle ,
\end{equation}
depends on coordinates $x$, $x'$ and on the lateral distance
$y=\vert {\bf y}-{\bf r}'\vert$, $S(x,x';y)$.
The asymptotic large-$y$ behavior is of the form
\begin{equation} \label{asym-charge-charge}
S(x,x';y) \mathop{\simeq}_{y\to\infty} \frac{F(x,x')}{y^d} .
\end{equation}
For the previous one-component system of particles with charge $-e$,
$F(x,x')$ is related to $f(x,x')$ by $F(x,x') = e^2 f(x,x')$.
The counterpart of the one-component sum rule (\ref{sumruled}) is
\begin{equation} \label{sumruleD}
\int_0^{\infty} {\rm d}x \int_0^{\infty} {\rm d}x'\, F(x,x') 
= - \frac{2}{\beta s_d^2} .
\end{equation} 

According to Blume et al. \cite{Blum81}, the many-component generalization
of Eq. (\ref{Blumd}) reads as
\begin{equation} \label{BlumD}
\frac{\partial \rho(x)}{\partial (e\sigma)} = - s_d \beta
\int_{-\infty}^{\infty} {\rm d}y \int_0^{\infty} {\rm d}x'\, (x'-x) S(x,x';y) .
\end{equation}
The many-component generalization of the relation (\ref{Uf}) reads
\cite{Jancovici01}
\begin{equation} \label{UF} 
\int_{-\infty}^{\infty} {\rm d}y \int_0^{\infty} {\rm d}x'\, (x'-x) S(x,x';y)
= \frac{s_d}{2} \int_0^{\infty} {\rm d}x'\, F(x,x') ,
\end{equation}
so that
\begin{equation} \label{BlumDD}
\frac{\partial \rho(x)}{\partial (e\sigma)} = - \frac{s_d^2 \beta}{2}
\int_0^{\infty} {\rm d}x'\, F(x,x') .
\end{equation}

Let us presuppose that the $F$-function factorizes as
\begin{equation}
F(x,x') = - G(x) G(x') .
\end{equation}
Taking into account the sum rule (\ref{sumruleD}) and Eq. (\ref{BlumDD}),
we find the direct local relation between the $G$-function and the 
charge profile: 
\begin{equation} \label{rr}
G(x) = \sqrt{\frac{2}{\beta}} \frac{1}{s_d} 
\left\vert \frac{\partial \rho(x)}{\partial (e\sigma)} \right\vert .
\end{equation}
Note that $G(x)$ is determined up to an irrelevant sign; for simplicity,
we have chosen $G(x)>0$.

\subsection{Exactly solvable cases}
In the Debye-H\"uckel high-temperature limit \cite{Jancovici82a},
the charge density profile takes the form
\begin{equation}
\rho(x,\sigma) = \rho(x,\sigma=0) - e \sigma \kappa {\rm e}^{-\kappa x} ,
\end{equation}
where $\kappa=\sqrt{s_d\beta e^2 \sum_{\alpha} q_{\alpha}^2 n_{\alpha}}$ 
is the multi-component inverse Debye length.
The asymptotic amplitude function $F(x,x')$ was found in the form 
\begin{equation}
F(x,x') = - \frac{2\kappa^2}{\beta s_d^2} {\rm e}^{-\kappa (x+x')} ,
\end{equation}
implying
\begin{equation}
G(x) = \sqrt{\frac{2}{\beta}} \frac{\kappa}{s_d} {\rm e}^{-\kappa x} .
\end{equation}
Since $\partial \rho(x)/\partial(\sigma e) = - \kappa {\rm e}^{-\kappa x}$,
it is trivial to verify that this $G$-function satisfies Eq. (\ref{rr}).

Another exactly solvable case is the 2D two-component plasma (Coulomb gas)
of $\pm e$ charges at coupling $\Gamma=2$ \cite{Cornu89,Jancovici92}.
The density profiles of $\pm e$ particles read as
\begin{equation}
n_{\pm}(x,\sigma) = n_{\pm}(x,\sigma=0) + \frac{m^2}{2\pi}
\int_0^{\mp 2\pi\sigma} \frac{{\rm d}t}{\sqrt{m^2+t^2}-t} 
{\rm e}^{-2\sqrt{m^2+t^2} x} ,
\end{equation}
where $m$ is a rescaled fugacity which has dimension of an inverse length.
Since $n_+(x,\sigma=0) = n_-(x,\sigma=0)$, the charge density
$\rho(x) = e \left[ n_+(x) - n_-(x) \right]$ results in
\begin{equation}
\rho(x) = - \frac{e}{\pi} \int_0^{2\pi\sigma} \sqrt{m^2+t^2} 
{\rm e}^{-2\sqrt{m^2+t^2} x} . 
\end{equation}
Introducing the variable $k_0 = \sqrt{m^2+(2\pi\sigma)^2}$, we obtain that
\begin{equation}
\frac{\partial \rho(x)}{\partial (e\sigma)} = - 2 k_0 {\rm e}^{-2k_0 x} .
\end{equation}
Simultaneously, it holds \cite{Cornu89} 
\begin{equation}
F(x,x') = - \frac{k_0^2 e^2}{\pi^2} {\rm e}^{-2k_0 (x+x')} , \qquad
G(x) = - \frac{k_0 e}{\pi} {\rm e}^{-2k_0 x} .
\end{equation}
Taking $\beta e^2 = 2$, Eq. (\ref{rr}) is readily shown to be satisfied. 

\renewcommand{\theequation}{6.\arabic{equation}}
\setcounter{equation}{0}

\section{Conclusion}
This paper was motivated by the previous one \cite{Samaj15} where,
using the technique of anticommuting variables for a 2D model of the charged
wall with counter-ions only, a new relation was found between the amplitude 
function $f(x,x')$ (with $x'=0$) of the asymptotic decay of two-body densities 
along the wall and the particle density profile $n(x)$, see Eq. (\ref{rov3}).
Here in Sect. 2, using the M\"obius conformal transformation of particle
coordinates on the level of one-body density for the same model, the
complementary relation (\ref{rov2}) was derived.
The combination of the two exact relations enabled us to prove 
the factorization property $f(x,x') = - g(x) g(x')$ and to express $g(x)$ 
in terms of the density profile.

For more-complicated many-component Coulomb fluids in any dimension,
it is necessary to concentrate on the charge-charge structure function
(\ref{charge-charge}) with the asymptotic behavior (\ref{asym-charge-charge})
and to look on the relation between the amplitude function $F(x,x')$ and
the charge density profile $\rho(x)$.
In all exactly solvable cases which are available in the high-temperature 
limit and at the 2D free-fermion coupling, the amplitude function 
$F(x,x')$ factorizes.
There is no proof of the factorization property of the amplitude function 
at any temperature.
In general, the statistical independence of two particles at asymptotically 
large distances is reflected by the nullity of the truncated correlation 
functions.
In our semi-infinite problem, the distance between two particles goes
to infinity along the wall, $y\to\infty$, but the distances of the particles 
from the wall $x,x'$ are finite. 
One can intuitively argue that the limit $y\to\infty$ automatically decouples 
the subspaces $x$ and $x'$ which is behind the factorization property of 
the amplitude function. 
Presupposing $F(x,x') = - G(x) G(x')$ for any Coulomb fluid, the combination 
of two sum rules (\ref{BlumD}) and (\ref{UF}) permits us to express $G(x)$ 
in terms of the charge density profile $\rho(x)$, see Eq. (\ref{rr}).

As concerns future perspective, it would be desirable to find simplified
models or new methods to prove the factorization property of the amplitude 
function for more general Coulomb fluids.
A better comprehension of the form of the amplitude function might clarify
the form of the dielectric susceptibility tensor for an arbitrarily
shaped domain.
     
\begin{acknowledgements}
The support received from Grant VEGA No. 2/0015/15 is acknowledged. 
\end{acknowledgements}


\begin{thebibliography}{10}

\bibitem{Andelman} Andelman, D.:
Introduction to electrostatics in soft and biological matter. 
In: Poon, W.C.K., Andelman, D. (eds.) Soft Condensed Matter Physics
in Molecular and Cell Biology, vol. 6. Taylor \& Francis, New York, (2006)

\bibitem{Attard88} Attard, P., Mitchell, D.J., Ninham, B.W.:
Beyond Poisson-Boltzmann: Images and correlations in the electric double
layer. I. Counterions only.
J. Chem. Phys. {\bf 88}, 4987--4996  (1988)

\bibitem{Attard96} Attard, Ph.:
Electrolytes and the electric double layer.
Adv. Chem. Phys. {\bf XCII}, 1--159 (1996)

\bibitem{Blum81} Blum, L., Henderson, D., Lebowitz, J.L., Gruber, Ch., Martin, 
Ph.A.: A sum rule for an inhomogeneous electrolyte.
J. Chem. Phys. {\bf 75}, 5974--5975 (1981) 

\bibitem{Boroudjerdi05} Boroudjerdi, H., Kim, Y.-W., Naji, A.,Netz, R.R.,
Schlagberger, X., Serr, A.:
Statics and dynamics of strongly charged soft matter.
Phys. Rep. {\bf 416}, 129--199 (2005)

\bibitem{Choquard86} Choquard, Ph., Piller, B., Rentsch, R.: 
On the dielectric susceptibility of classical Coulomb systems.
J. Stat. Phys. {\bf 43}, 197--205 (1986)

\bibitem{Choquard87} Choquard, Ph., Piller, B., Rentsch, R.: 
On the dielectric susceptibility of classical Coulomb systems. II.
J. Stat. Phys. {\bf 46}, 599--633 (1987)

\bibitem{Choquard89} Choquard, Ph., Piller, B., Rentsch, R., Vieillefosse, P.: 
Surface properties of finite classical Coulomb systems: Debye-H\"uckel 
approximation and computer simulations. 
J. Stat. Phys. {\bf 55}, 1185--1262 (1989)

\bibitem{Cornu89} Cornu, F., Jancovici, B.:
The electrical double layer: A solvable model.
J. Chem. Phys. {\bf 90}, 2444-2452 (1989)

\bibitem{Hansen00} Hansen, J.P., L\"owen, H.:
Effective interactions between electric double layers.
Annu. Rev. Phys. Chem. {\bf 51}, 209--242  (2000) 

\bibitem{Jancovici82a} Jancovici, B.:
Classical Coulomb systems near a plane wall. I.
J. Stat. Phys. {\bf 28}, 43--65 (1982)

\bibitem{Jancovici82b} Jancovici, B.:
Classical Coulomb systems near a plane wall. II.
J. Stat. Phys. {\bf 29}, 263--280 (1982)

\bibitem{Jancovici84} Jancovici, B.:
Surface properties of a classical two-dimensional one-component plasma: 
Exact results.
J. Stat. Phys. {\bf 34}, 803--815 (1984)

\bibitem{Jancovici92} Jancovici, B.:
Inhomogeneous two-dimensional plasmas.
In: Henderson. D. (ed.) Inhomogeneous Fluids, pp. 201-237, Dekker, 
New York (1992)

\bibitem{Jancovici95} Jancovici, B.:
Classical Coulomb systems: Screening and correlations revisited.
J. Stat. Phys. {\bf 80}, 445--459 (1995)

\bibitem{Jancovici01} Jancovici, B., \v{S}amaj, L.:
Charge correlations in a Coulomb system along a plane wall:
A relation between asymptotic behavior and dipole moment.
J. Stat. Phys. {\bf 105}, 193--209 (2001)

\bibitem{Jancovici02} Jancovici, B.:
Surface correlations for two-dimensional Coulomb fluids in a disc.
J. Phys.: Condens. Matter {\bf 14}, 9121--9132 (2002)

\bibitem{Mallarino13} Mallarino, J.P., T\'ellez, G., Trizac, E.:
Counter-ion density profile around charged cylinders: the strong-coupling
needle limit.
J. Phys. Chem. B {\bf 117}, 12702--12716 (2013)

\bibitem{Mallarino15} Mallarino, J.P., T\'ellez:
Counter-ion density profile around a charged disc: from the weak to the strong
association regime.
Phys. Rev. E {\bf 91}, 062140 (2015)

\bibitem{Martin88} Martin, Ph.A.:
Sum rules in charged fluids.
Rev. Mod. Phys. {\bf 60}, 1075--1127 (1988)

\bibitem{Messina09} Messina, R.:
Electrostatics in soft matter.
J. Phys.: Condens. Matter {\bf 21}, 113102 (2009)

\bibitem{Netz00} Netz, R.R., Orland, H.:
Beyond Poisson-Boltzmann: Fluctuation effects and correlation functions.
Eur. Phys. J. E {\bf 1}, 203--214 (2000)

\bibitem{Netz01} Netz, R.R.:
Electrostatics of counter-ions at and between planar charged walls: from
Poisson-Boltzmann to the strong-coupling theory.
Eur. Phys. J. E {\bf 5}, 557--574 (2001)

\bibitem{Podgornik90} Podgornik, R.:
An analytic treatment of the first-order correction to the Poisson-Boltzmann
interaction free energy in the case of counter-ion only Coulomb fluid.
J. Phys. A: Math. Gen. {\bf 23}, 275--284 (1990)

\bibitem{Samaj95} \v{S}amaj, L., Percus, J.K.:
A functional relation among the pair correlations of the two-dimensional
one-component plasma.
J. Stat. Phys. {\bf 80}, 811--824 (1995)

\bibitem{Samaj00} \v{S}amaj, L.:
Microscopic calculation of the dielectric susceptibility tensor for 
Coulomb fluids.
J. Stat. Phys. {\bf 100}, 949--967 (2000)

\bibitem{Samaj04} \v{S}amaj, L.:
Is the two-dimensional one-component plasma exactly solvable?
J. Stat. Phys. {\bf 117}, 131--158 (2004)

\bibitem{Samaj10} \v{S}amaj, L., Jancovici, B.:
Charge and current sum rules in quantum media coupled to radiation II
J. Stat. Phys. {\bf 139}, 432--453 (2010)

\bibitem{Samaj11a} \v{S}amaj, L., Trizac, E.:
Counterions at highly charged interfaces: From one plate to like-charge
attraction.
Phys. Rev. Lett. {\bf 106}, 078301 (2011)

\bibitem{Samaj11b} \v{S}amaj, L., Trizac, E.:
Counter-ions at charged walls: Two-dimensional systems.
Eur. Phys. J. E {\bf 34}, 20 (2011)

\bibitem{Samaj13} \v{S}amaj, L.:
Counter-ions at single charged wall: Sum rules.
Eur. Phys. J. E {\bf 36}, 100 (2013)

\bibitem{Samaj14} \v{S}amaj, L., Trizac, E.:
Counter-ions between or at asymmetrically charged walls: 2D free-fermion point.
J. Stat. Phys. {\bf 156}, 932--947 (2014)

\bibitem{Samaj15} \v{S}amaj, L.:
Counter-ions near a charged wall: Exact results for disc and planar geometries.
J. Stat. Phys. {\bf 161}, 227--249 (2015)

\bibitem{Tutschke} Tutschke, W., Vasudeva, H.L.:
An introduction to complex analysis: Classical and modern approaches. 
CRC Press, London (2004).

\bibitem{Usenko79} Usenko, A.S., Yakimenko, I.P.:
Interaction energy of stationary charges in a bounded plasma.
Sov. Tech. Phys. Lett. {\bf 5}, 549--550 (1979) 

\end{thebibliography}
\end{document}